\documentclass{article}
\usepackage{color}
\usepackage[utf8]{inputenc}
\usepackage[T1]{fontenc}
\usepackage{cite}
\usepackage{fixltx2e}
\usepackage{graphicx}
\usepackage{url}
\usepackage{multirow}
\usepackage{amsmath}
\usepackage[standard]{ntheorem}
\usepackage{subfigure}
\usepackage[english]{babel}
\usepackage[T1]{fontenc}
\usepackage{textcomp}
\usepackage{lmodern}
\usepackage[utf8]{inputenc}
\usepackage{pxfonts}
\usepackage{dsfont}
\usepackage{stmaryrd}
\usepackage{listings}
\begin{document}

\title{Computational investigations of folded self-avoiding walks related to protein folding}
\author{Jacques M. Bahi, Christophe Guyeux, Kamel Mazouzi, and Laurent Philippe\thanks{Authors in alphabetic order}\\
  Computer science laboratory DISC,\\ FEMTO-ST Institute, UMR 6174 CNRS\\
 University of Franche-Comt\'{e}, Besan\c con, France\\
 \textit{\{jacques.bahi, christophe.guyeux, kamel.mazouzi, laurent.philippe\}@femto-st.fr}}
 
\maketitle

\begin{abstract}
Various subsets of self-avoiding walks naturally appear when investigating
existing methods designed to predict the 3D conformation of a 
protein of interest. Two such subsets, namely the folded and the
unfoldable self-avoiding walks, are studied computationally in this
article. We show that these two sets are equal and correspond to the
whole $n$-step self-avoiding walks for $n\leqslant 14$, but that
they are different for numerous $n \geqslant 108$, which are common
protein lengths. Concrete counterexamples are provided and the 
computational methods used to discover them are completely detailed.
A tool for studying these subsets of walks related to both pivot moves
and proteins conformations is finally presented.
\end{abstract}

\section{Introduction}

Self-avoiding walks (SAWs) have been studied over decades for the
extent and difficulty of the mathematical problems they 
provide~\cite{Sokal88,Bousquet11,Beaton12},
and for their various contexts of application in physics, chemistry,
and biology~\cite{Gordon11,Gennes72,Flory49}. 
Among other things, they are used to model polymers such
as DNA, RNAs, and proteins. Numerous protein structure prediction 
(PSP) software iterate on self-avoiding walk subsets, often not 
clearly defined, of various lattices, in such a way that the last
produced SAW $S$ has the length of the targeted protein $P$ and,
when labeling $S$ with the amino acids of $P$, $S$ is (one of) the
best solution(s) according to a scoring function that associates
a value to a 2D or 3D conformation (depending on physical properties
of the conformation as hydrophobic neighboring residues, etc.).

In previous studies~\cite{guyeux:hal-00795127,bgcs11:ij,bgc11:ip}, 
authors of this manuscript have investigated some protein folding models of dynamics. They have shown that
the possible sets of conformations reachable by these numerous PSP
software are not equal, which raises severe questionings on what is
indeed really predicted by such software. In particular, they 
have shown that software that iteratively stretch the conformation
from one amino acid until a self-avoiding walk having the length
$n$ of the protein, can reach all the $n$-step SAWs $\mathfrak{G}_n$.
Contrarily, the ones that iterate $\pm90^\circ$ pivot moves on the 
$n$-step straight line can only reach what they called the subset
of folded self-avoiding walks $fSAW(n)$. It has been clearly 
established that, for some well-defined small $n$`s, 
$fSAW(n) \neq \mathfrak{G}_n$. After having obtained this result,
the authors' intention was then to investigate more deeply these
new kind of self-avoiding walks and other related subsets of walks
they called unfolded SAWs, and to determine consequences of these
investigations regarding the protein structure prediction problem.

This article is the third of a series of three researches we 
publish in that field. In~\cite{articleGeneral} we provide a general
presentation of folded and unfoldable SAWs, and the collection of 
results we have obtained on these objects using both theoretical
and computational approaches. Article~\cite{articleTheoreme}
focuses more specifically on the mathematical study of theses 
subsets of self-avoiding walks, by proving in particular that
the number of unfolded SAWs is infinite. This article, for its part,
presents our computational investigations in detail.

After having recalled in the next section the basis of self-avoiding walk, of folded
SAWs obtained by iterating pivot moves on the straight line, and
of unfoldable SAWs on which no pivot move can be applied without
breaking the self-avoiding property, we explain in 
Section~\ref{nbofFoldedsaws} how the number of
folded self-avoiding walks has been computed and how we checked the
unfoldable property in practice. The various methods that have been implemented
to find the shortest currently known unfoldable SAW are presented
too in this section. Then, in Section~\ref{sec:heuristics}, some 
heuristics that could be determinant in further studies concerning
these subsets of walks are introduced. The next section contains
the last contribution of this research work: a free software 
realized to facilitate the study of folded and unfoldable SAWs.
This document ends by a conclusion section, in which all these
contributions are summarized and intended future work is proposed.

\section{Presentation of Folded Self-Avoiding Walks}

We recall in this section various notions and properties of self-avoiding
walks and of some of theirs folded subsets. Authors that would investigate
more deeply these walks are referred to~\cite{Madras93,Gordon11,Hughes95}
for the SAWs in general, and to~\cite{articleGeneral,articleTheoreme} for
the folded case.

\subsection{Definitions and Terminologies}

Let $\mathds{N}$ be the set of all natural numbers,
$\mathds{N}^*=\{1,2,\hdots\}$ the set of all positive integers, and
for $a,b\in \mathds{N}$, $a\neq b$, the notation $\llbracket
a,b\rrbracket$ stands for the set $\{a, a+1, \hdots, b-1, b\}$.  $|x|$
stands for the Euclidean norm of any vector $x\in \mathds{Z}^d,
d\geqslant 1$, whereas $x_1, \hdots, x_n$ are the $n$ coordinates of
$x$.  The $n-$th term of a sequence $s$ is denoted by $s(n)$.
Finally, $\sharp X$ is the cardinality of a finite set $X$.
Using this material, self-avoiding walk can be defined 
as follows~\cite{Madras93,Gordon11,Hughes95}.

\begin{figure}
\centering
\includegraphics[scale=0.30]{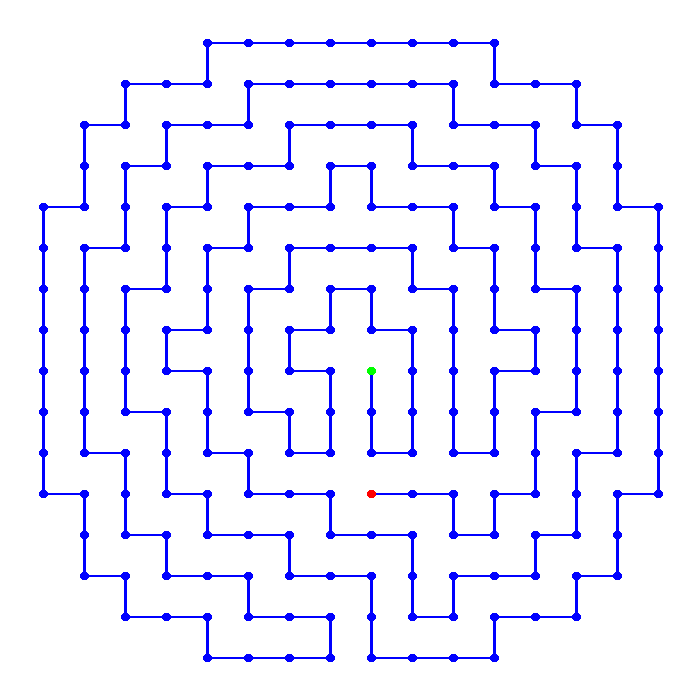}
\caption{The first SAW shown to be not connected
to any other SAW by 90° rotations (Madras and Sokal,~\cite{Sokal88}).}
\label{SokalnrSAW}
\end{figure}

\begin{definition}[Self-Avoiding Walk]
  Let $d\geqslant 1$. A $n-$step \emph{self-avoiding walk} from
  $x\in\mathds{Z}^d$ to $y\in\mathds{Z}^d$ is a map $w:\llbracket
  0,n\rrbracket \rightarrow \mathds{Z}^d$ with:
\begin{itemize}
\item $w(0)=x$ and $w(n)=y$,
\item $|w(i+1)-w(i)|=1$, 
\item $\forall i,j \in \llbracket 0,n\rrbracket$, $i\neq j \Rightarrow
  w(i)\neq w(j)$ (self-avoiding property).
\end{itemize}
\end{definition}

\subsection{Notations}

In absolute encoding~\cite{Hoque09,Backofen99algorithmicapproach}, a $n$-step walk $w=w(0), ..., w(n)\in \left(\mathds{Z}^2\right)^{n+1}$ with $w(0)=(0,0)$ is a sequence 
$s=s(0), ..., s(n-1)$ of elements belonging into $\mathds{Z}/4\mathds{Z}$,
such that:
\begin{itemize}
\item $s(i)=0$ if and only if $w(i+1)_1=w(i)_1+1$ and $w(i+1)_2=w(i)_2$, that is, $w(i+1)$ is at the East of $w(i)$.
\item $s(i)=1$ if and only if $w(i+1)_1=w(i)_1$ and $w(i+1)_2=w(i)_2-1$: $w(i+1)$ is at the South of $w(i)$.
\item $s(i)=2$ if and only if $w(i+1)_1=w(i)_1-1$ and $w(i+1)_2=w(i)_2$, meaning that $w(i+1)$ is at the West of $w(i)$.
\item Finally, $s(i)=3$ if and only if $w(i+1)_1=w(i)_1$ and $w(i+1)_2=w(i)_2+1$ ($w(i+1)$ is at the North of $w(i)$).
\end{itemize}

Let us now define the following functions~\cite{guyeux:hal-00795127}.

\begin{definition}
The \emph{anticlockwise fold function} is the function $f: \mathds{Z}/4\mathds{Z} \longrightarrow \mathds{Z}/4\mathds{Z}$ defined by $f(x) = x-1 ~(\textrm{mod}~ 4)$ and the clockwise fold function is $f^{-1}(x) = x+1 ~(\textrm{mod}~ 4)$.
\end{definition}

Using the absolute encoding sequence $s$ of a $n-$step SAW $w$ that starts from the origin of the square lattice, a pivot move of $+90$° on $w(k)$, $k<n$, simply 
consists to transform $s$ into $s(0),\hdots,s(k-1),f(s(k)),\hdots,f(s(n))$. Similarly, a pivot move of $-90$° consists to apply $f^{-1}$
to the queue of the absolute encoding sequence, like
in Figure~\ref{ex1}.

\subsection{A graph structure for SAWs folding process}

We can now introduce a graph structure describing well the iterations of
$\pm 90$° pivot moves on a given self-avoiding walk.

\begin{figure}
\centering
\includegraphics[scale=0.35]{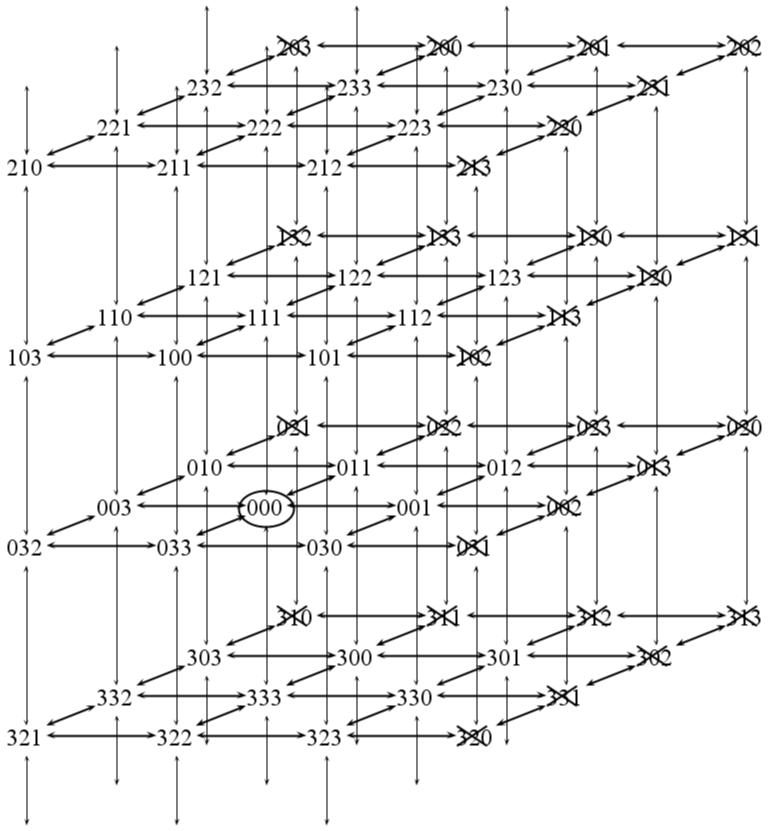}
\caption{The digraph $\mathfrak{G}(3)$}
\label{digraph}
\end{figure}

Given $n \in \mathds{N}^*$, the graph $\mathfrak{G}_n$, formerly introduced in~\cite{guyeux:hal-00795127}, is defined as follows:
\begin{itemize}
\item its vertices are the $n-$step self-avoiding walks, described in absolute 
encoding;
\item there is an edge between two vertices $s_i$, $s_j$ if and only if $s_j$ can be obtained by one pivot move of $\pm 90$° on $s_i$, that is, if there exists 
$k \in \llbracket 0, n-1\rrbracket$ s.t.:
\begin{itemize}
\item either $s_j(0),\hdots,s_j(k-1),f(s_j(k)),\hdots,f(s_j(n)) = s_i$
\item or $s_j(0),\hdots,s_j(k-1),f^{-1}(s_j(k)),\hdots,f^{-1}(s_j(n)) = s_i$.
\end{itemize}
\end{itemize}
Such a digraph is depicted in Figure~\ref{digraph}. The circled vertex is the
straight line whereas strikeout vertices are walks that are not self-avoiding.
Depending on the context, and for the sake of simplicity, $\mathfrak{G}_n$ will 
also refers to the set of SAWs in $\mathfrak{G}_n$ (\emph{i.e.}, its vertices).

\begin{figure}[h]
\centering
\subfigure[000111]{\includegraphics[scale=0.2]{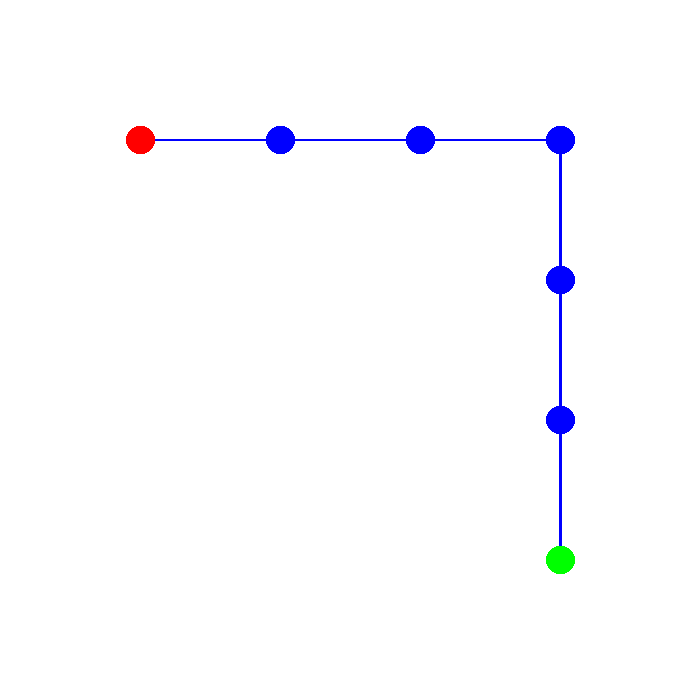}}\hspace{2cm}\subfigure[$001222=00f^{-1}(0)f^{-1}(1)f^{-1}(1)f^{-1}(1)$]{\includegraphics[scale=0.2]{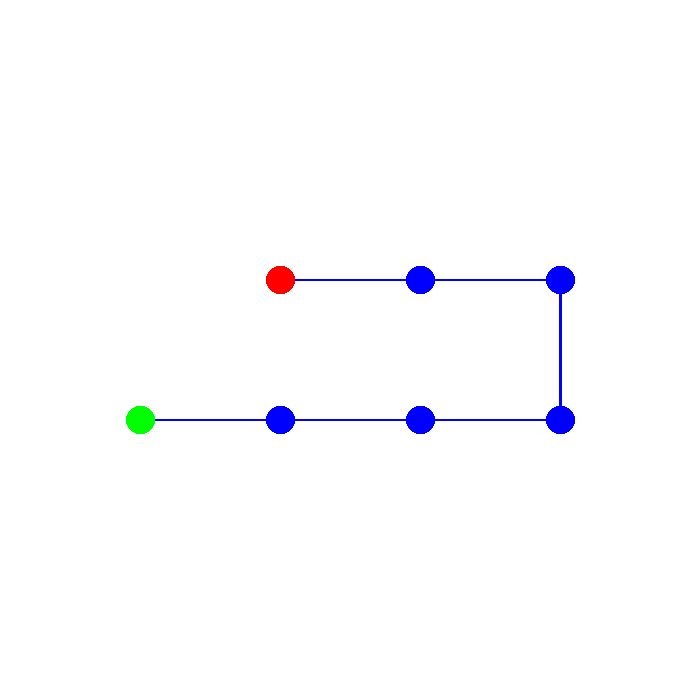}}
\caption{Effects of the clockwise fold function applied on the four last components of an absolute encoding.}
\label{ex1}
\end{figure}

Using this graph, the folded SAWs introduced in the previous section can be
redefined more rigorously.
\begin{definition}
$fSAW_n$ is the connected component of the straight line 
$00\hdots 0$ ($n$ times) in $\mathfrak{G}_n$, whereas $\mathcal{S}_n$ is 
constituted by all the vertices of $\mathfrak{G}_n$.
\end{definition}

\begin{figure}
\centering
\subfigure{\includegraphics[scale=0.23]{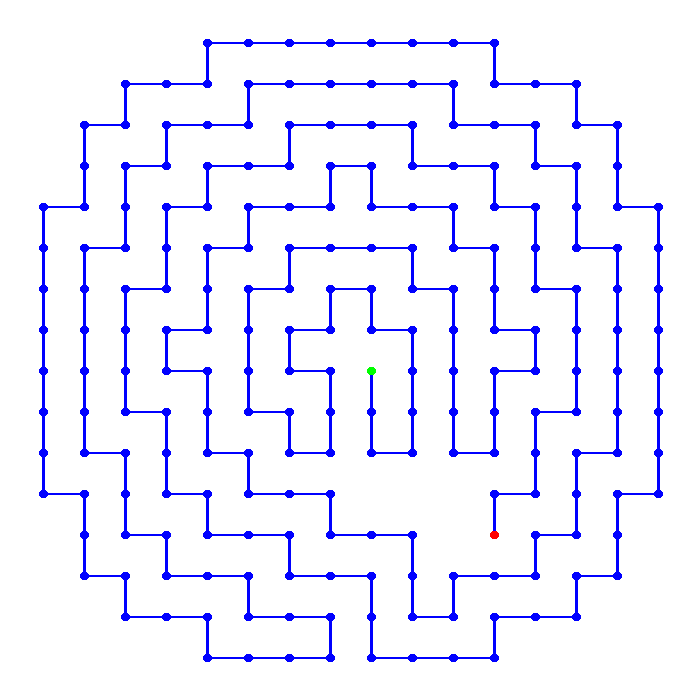}} \hspace{0.2cm} \subfigure{\includegraphics[scale=0.23]{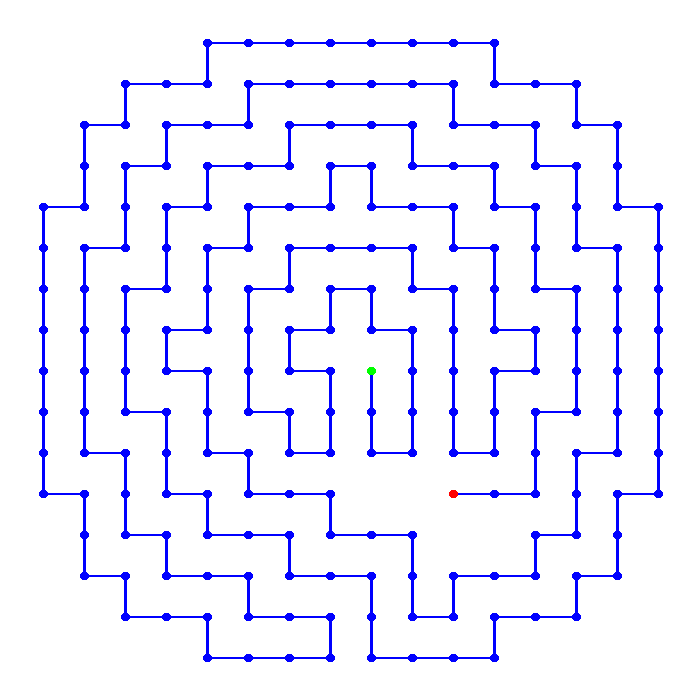}} 
\caption{A self-avoiding walk in $fSAW(219,2)$}
\label{saw3pasSaw4}
\end{figure}

The Figure~\ref{SokalnrSAW} shows that the connected component $fSAW(223)$ of
the straight line in
$\mathfrak{G}_{223}$ is not equal to the whole graph: $\mathfrak{G}_{223}$
is not connected. More precisely, this graph has a connected component of size 1:
it is unfoldable whereas SAW of Fig.~\ref{saw3pasSaw4} can be folded
exactly once. Indeed, to be in the same connected component is an
equivalence relation $\mathcal{R}_n$ on $\mathfrak{G}_n, \forall n \in \mathds{N}^*$, and two SAWs $w$, $w'$ are considered equivalent (with respect to this equivalence
relation) if and only if there is a way to fold $w$ into $w'$ such that all
the intermediate walks are self-avoiding. When existing, such a way is not
necessarily unique.

These remarks lead to the following definitions.

\begin{definition}
Let $n\in\mathds{N}^*$ and $w \in \mathcal{S}_n$. We say that:
\begin{itemize}
\item $w$ is \emph{unfoldable} if its equivalence class, with
  respect to $\mathcal{R}_n$, is of size 1;
\item $w$ \emph{is a folded self-avoiding walk} if its equivalence
  class contains the $n-$step straight walk $000\hdots 0$ ($n-1$
  times);
\item $w$ \emph{can be folded $k$ times} if  a simple path of
  length $k$ exists between $w$ and another vertex in the same connected
  component of $w$.
\end{itemize}
Moreover, we introduce the following sets:
\begin{itemize}
\item $fSAW(n)$ is the equivalence class of the $n-$step straight walk, or the set of all folded SAWs.
\item $fSAW(n,k)$ is the set of equivalence classes of size $k$ in $(\mathfrak{G}_n,\mathcal{R}_n)$.
\item $USAW(n)$ is the set of equivalence classes of size 1
  $(\mathfrak{G}_n,\mathcal{R}_n)$, that is, the set of unfoldable walks.
\item $f^1SAW(n)$ is the complement of $USAW(n)$ in
  $\mathfrak{G}_n$. This is the set of SAWs on which we can apply at
  least one pivot move of $\pm 90^{\circ}$.
\end{itemize}
\end{definition}

\begin{example}
Fig.~\ref{ex1} shows the two elements of a class belonging into $fSAW(219,2)$
whereas Fig.~\ref{SokalnrSAW} is an element of $USAW(223)$.
\end{example}

\section{Computing the number of folded self-avoiding walks}
\label{nbofFoldedsaws}

\subsection{The number $\sharp\mathfrak{G}_n$ of all possible $s$-step SAWs}

To recall in detail how the number $\sharp\mathfrak{G}_n$ of all 
possible SAWs has been computed until $n=71$ is not the goal of
this article. Let us just remark that,
in~\cite{Conway1993}, Conway \emph{et al.} have presented an algebraic
technique for enumerating self-avoiding walks on a rectangular lattice,
by rewriting the generating function for SAWs on this lattice 
using 5 irreducible components, whose generating functions are easier to 
calculate (see Formula 2.17 of~\cite{Conway1993}). 
These irreducible components are obtained by considering projections of
walks onto the $y$ axis, and classify irreducible segments by the number
of $y$ bonds they span.
The obtained cardinality of $\mathfrak{G}_n$, 
is recalled in Table~\ref{composante connexe} for $	n\leqslant 31$. 
For further detail, the reader is referred to~\cite{Conway1993}.

Authors of this research work
 have tried to adapt this very interesting approach 
by searching a way to decompose the generating function of 
folded self-avoiding walks in other generating functions easier to calculate.
However the way to take into account pivot moves that define 
such folded SAWs has not yet been discovered. 
This is why the algebraic method has been abandoned after unsuccessful
attempts to the benefit to a constructive brute force approach detailed
in the following sections.

At each of our investigations, a draft program in Python language has been released first,
to test rapidly the correctness of the approach. This code as been
translated in an optimized C program and deployed in the supercomputer
facilities of the Mésocentre de calcul de Franche-Comté.
For readability and compactness, only the python programs are presented 
thereafter.

\subsection{Preliminaries}

Python function called \emph{walks} (Listing~\ref{AllConformations}) 
produces the list of all possible $n$-step walks as follows: the 
$n$-step walks are the walks of lenght $n-1$ with 0, 1, 2, or
3 added to their tails (recursive call). The return is a list of
walks, that is, a list of integers lists.

\begin{figure}[h]
\begin{lstlisting}[language=Python,frame=single,breaklines=true,numbers=left,basicstyle=\scriptsize]
def walks(n):
    if n==1:
        return [[0]]
    else:
        L = []
        for k in walks(n-1):
            for i in range(4):
                L.append(k+[i])
        return L
\end{lstlisting}
\caption{Obtaining all the walks}
\label{AllConformations}
\end{figure}

To obtain the walks belonging into $\mathfrak{G}_n$, we 
first introduce the function \emph{points} which aims is
to produce the list of points (two coordinates) of the
square lattice that corresponds to a given walk $C$.
Programs that encodes a walk described as a list of points in its
absolute encoding is provided in Listing~\ref{encodes} with
its \emph{decodes} associated function.

\begin{figure}[h]
\begin{lstlisting}[language=Python,frame=single,breaklines=true,numbers=left,basicstyle=\scriptsize]
def encodes(points):
    L=[]
    for k in range(1,len(points)):
        if points[k-1][0]==points[k][0]:
            if points[k-1][1] < points[k][1]:
                L.append(3)
            else:
                L.append(1)
        else:
            if points[k-1][0] < points[k][0]:
                L.append(0)
            else:
                L.append(2)
    return L

def decodes(C):
    L = [(0,0)]
    for c in C:
        P = L[-1]
        if c == 0:   L.append((P[0]+1,P[1]))
        elif c == 1: L.append((P[0],P[1]-1))
        elif c == 2: L.append((P[0]-1,P[1]))
        elif c == 3: L.append((P[0],P[1]+1))
    return L
\end{lstlisting}
\caption{Encoding and decoding walks}
\label{encodes}
\end{figure}

Function \emph{is\_saw} returns a Boolean: it is true if and
only if the walk $C$ satisfies the self-avoiding property. To do so, the
list of its points in the lattice (its support) is produced, and it is
regarded whether this list contains twice a same point (in
other words, if the support has the same size than 
the list of points).

Finally, \emph{saws} function produces a generator. It
returns the next SAW at each call of the 
\emph{next} method on the generator. To do so, an exhaustive
iteration of the list produced by \emph{walks} is
realized, and the \emph{is\_saw} function is applied to each 
element of this list, to test if this walk is self-avoiding.

\begin{figure}[h]
\begin{lstlisting}[language=Python,frame=single,breaklines=true,numbers=left,basicstyle=\scriptsize]
def points(C):
    L = [(0,0)]
    for c in C:
        P = L[-1]
        if c == 0: L.append((P[0]+1,P[1]))
        elif c == 1: L.append((P[0],P[1]-1))
        elif c == 2: L.append((P[0]-1,P[1]))
        elif c == 3: L.append((P[0],P[1]+1))
    return L

def is_saw(C):
    L = points(C)
    return len(L) == len(list(set(L)))

def saws(n):
    for k in walks(n):
        if is_saw(k):
            yield k
\end{lstlisting}
\caption{Finding the self-avoiding walks}
\label{SAW1Conformations}
\end{figure}

These functions have been written and optimized in C language.

\subsection{A backtracking method to discover unfoldable walks}

All the possible pivot moves (either in the clockwise direction, or in the anticlockwise) are checked to determine if a self-avoiding walk is unfoldable.
The \emph{fold} function tests, considering a given walk, a pivot
move on residue number \emph{position} following the given \emph{direction} (+1 or
-1, if clockwise or not). Function \emph{is\_unfoldable} applies the \emph{fold} function
to each residue of the candidate, and for the two possible directions. 
The function returns True if and only if no pivot move is possible.

\begin{figure}[h]
\begin{lstlisting}[language=Python,frame=single,breaklines=true,numbers=left,basicstyle=\scriptsize]
def fold(walk, position, direction):
    if position == 0:
        return walk
    new = []
    for k in range(len(walk)):
        if k<abs(position):
            new.append(walk[k])
        else:
            new.append((walk[k]+direction)%4)
    return decodes(new)

def is_unfoldable(saw):
    for k in range(1,len(saw)):
        if is_saw(fold(saw, k, -1)):
            return False
        elif is_saw(flod(saw, k, +1)):
            return False
    return True
\end{lstlisting}
\caption{Testing whether a self-avoiding walk is unfoldable}
\label{SAW3Conformations}
\end{figure}

Listing~\ref{backtrack} details how to enumerate walks in $f^1SAW(n)$
(the complement of $USAW(n)$) by constructing using a backtracking
method all the $k$-step walks for $k$ lower than a given threshold $n$,
and increasing a counter at each time the walk is not unfoldable
and of length $n$. The factor 4 at the end of the program is due
to the fact that we consider that the first step of these walks is 1
(South).

\begin{figure}[h]
\begin{lstlisting}[language=Python,frame=single,breaklines=true,numbers=left,basicstyle=\scriptsize]
def stretch(P,c):
    if c == 0:   return (P[0]+1,P[1])
    elif c == 1: return (P[0],P[1]-1)
    elif c == 2: return (P[0]-1,P[1])
    else :       return (P[0],P[1]+1)

def backtrack(w,n):
    global nb
    if len(w)>=n:
        return
    for a in range(4):
        u = stretch(w[-1],a)
        if u not in set(w):
            w1 = w+[u]
            if not is_unfoldable(encodes(w1)):
                if len(w1)==n-1:
                    nb += 1
                    # print ''.join([str(k) for k in w1])
            backtrack(w1,n)

nb=0
n=13
backtrack(decodes([1]),n+2)  
print n,4*nb
\end{lstlisting}
\caption{Backtracking method of $USAW(n)$}
\label{backtrack}
\end{figure}

Backtracking method of Listing~\ref{backtrack} has been translated
in C language, optimized, parallelized, and launched on the supercomputer facilities
(at each time, SAWs are separated over available processors 
using MPI routines) as follows.

A first version of the program, with the print line above turned as non commented,
has been launched with $N=14$ and $N=20$. By doing so, all the walks of $\mathfrak{G}_{N}$
that start in the East direction have been obtained (for $N=20$, they represent 
224424291 walks stored in 3 gigabytes of data). 
Then each of these $|\mathfrak{G}_{N}|/4$ self-avoiding walks has been the starting point
of another backtracking discovery until $n>N$, and the unfoldable property
of each of these $n$-step walk has been finally tested. 
This systematic approach as been successively launched until reaching $n=28$,
see Table~\ref{composante connexe}.
For $(N,n)=(20,28)$, 64 processors have been used during 70 hours in order to
test the unfoldable property of 2351378582244 $28$-step self-avoiding walks,
whereas no result has been obtained after 20 days of computation using
the same facilities with $(N,n)=(28,30)$.
We can summarize these results as follows.

\begin{proposition}
$\forall n \leqslant 28, f^1SAW(n) = \mathfrak{G}_n$.
\end{proposition}

\subsection{Investigating the $fSAW(n)$ set}

In the previous section, self-avoiding walks that can be
folded at least once have been enumerated. That is,
the cardinality of $f^1SAW(n)$ has been obtained for $n\leqslant 28$.
A breadth first search is now presented to show how
$\mathfrak{G}_n = fSAW(n)$ has been obtained for $n\leqslant 14$.

\begin{figure}[h]
\begin{lstlisting}[language=Python,frame=single,breaklines=true,numbers=left,basicstyle=\scriptsize]
from networkx import *

def fold(walk, position, direction):
    new = []
    for k in range(len(walk)):
        if k<abs(position):
            new.append(walk[k])
        else:
            new.append((walk[k]+direction)%4)
    return new

def toString(encoding_list):
    return "".join([str(k) for k in encoding_list])
\end{lstlisting}
\caption{Preliminaries for $fSAW(n)$ investigations}
\label{prelim}
\end{figure}

The graph structure used in the Python draft program has been provided by
the networkx library~\cite{networkx}, which is thus imported in the first
line of Listing~\ref{prelim}. 
The graph $G$ representing $\mathfrak{G}(n)$ is instantiated in the first line of Listing~\ref{mainExplore},
and the three following lines of this listing adds the node of the $n$-step 
straight line, shown as a word of $n$ zeros. The association between relative
encoding of a $n$-step walk described as an integer list and words on the  
alphabet $\{0,1,2,3\}$ of length $n$ in the nodes of $G$ is provided by function 
\emph{toString} of Listing~\ref{prelim}.

\begin{figure}[h]
\begin{lstlisting}[language=Python,frame=single,breaklines=true,numbers=left,basicstyle=\scriptsize]
def explore(G,node):
    future_nodes = []
    for k in range(1,len(node)):
        for direction in [-1,1]:
            new = fold(node, k, direction)
            newString = toString(new)
            if not G.has_node(newString):
                if is_saw(new):
                    future_nodes.append(new)
                    G.add_edge(toString(node), newString)
            else:
                G.add_edge(toString(node), newString)
    return (G,future_nodes)
\end{lstlisting}
\caption{The breadth first search \emph{explore} function}
\label{BFS}
\end{figure}

Let us recall that there is an edge between two nodes in $G$ if and only if 
the SAW of the second node can be obtained by a pivot move on the first walk.
This pivot moves is realized on the integer lists of the relative encoding 
of the walk using the \emph{fold} function presented in the
previous section. This latter must be adapted a little, to match with the fact
that $G$ contains words (not lists of integers). This adaptation is given too
in Listing~\ref{prelim}.

Then the whole connected component of the straight line is constructed using
a breadth first search approach: at each iteration, the list of new walks 
that result from a fold on the last added nodes is obtained (with function
\emph{explore} of Listing~\ref{BFS}) and required connections are provided
between last added walks and new discovered ones.
The main issue in this approach is to prevent from visiting twice a given node,
which is verified with the \emph{G.has\_node} method.

\begin{figure}[h]
\begin{lstlisting}[language=Python,frame=single,breaklines=true,numbers=left,basicstyle=\scriptsize]
G = Graph()
n=6
nodes = [[0]*n]
G.add_node(toString(nodes[0]))

while nodes != []:
    new_nodes=[]
    for node in nodes:
        [G,x] = explore(G,node)
        for k in x:
            if k not in new_nodes:
                new_nodes.append(k)
    nodes = new_nodes

print(4*len(G))
\end{lstlisting}
\caption{Main program to compute the size of the connected component of the straight line.}
\label{mainExplore}
\end{figure}

More precisely, given a last added node, function \emph{explore} realizes an 
one depth exploration starting from this node, adds the new discovered nodes
to $G$ with related edges, and returns $G$ with the list of these new nodes $x$.
This $x$ is used in Listing~\ref{mainExplore} to constitute the next depth
of exploration.

\subsection{In search of the shortest unfoldable self-avoiding walks}

\begin{figure}[h]
\centering
\includegraphics[scale=0.3]{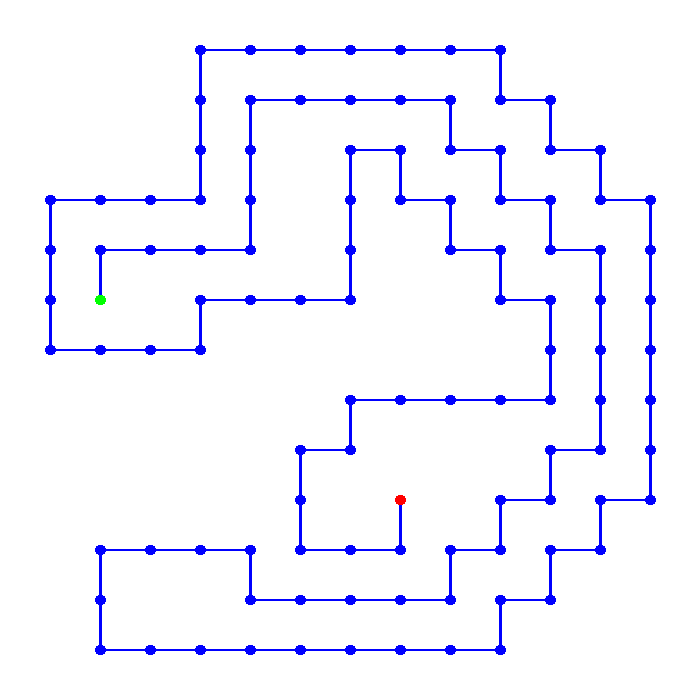}
\caption{Current smallest (107-step) unfoldable SAW}
\label{saw107}
\end{figure}

The current smallest unfoldable self-avoiding walk is an 107-step walk, as
depicted in Figure~\ref{saw107}. 
The production of this counterexample and the
systematic exploration of the connected component of the straight walk of
$\mathfrak{G}_n$ for small $n$'s presented previously allows us to claim that
(see
Table~\ref{composante connexe}):
\begin{proposition}
Let $\nu_n$ the smallest $n\geqslant 2$ such that $USAW(n) \neq \emptyset$.
Then $15\leqslant \nu_n\leqslant 107$. 
In other words, $\forall n \leqslant 14, fSAW(n) = \mathfrak{G}_n$, whereas $fSAW(107) \subsetneq \mathfrak{G}_{107}$.
\end{proposition}

\begin{figure}[h]
\centering
\subfigure[Pivot move at the border of a walk constituted by two intricately linked discs.]{\includegraphics[scale=0.6]{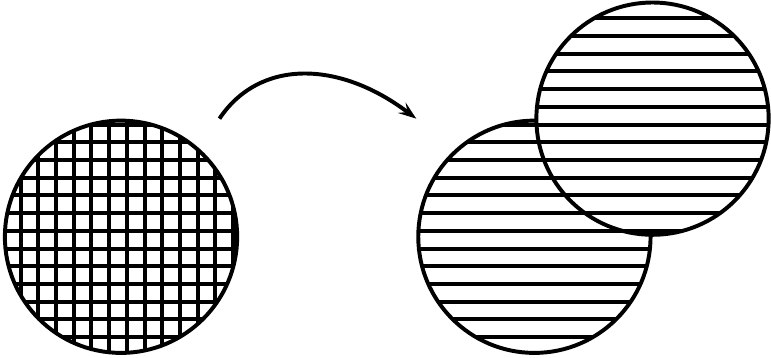}\label{pivotDunCercle}}
\hspace{1cm}\subfigure[First unfoldable SAW found, obtained by this approach.]{\includegraphics[scale=0.18]{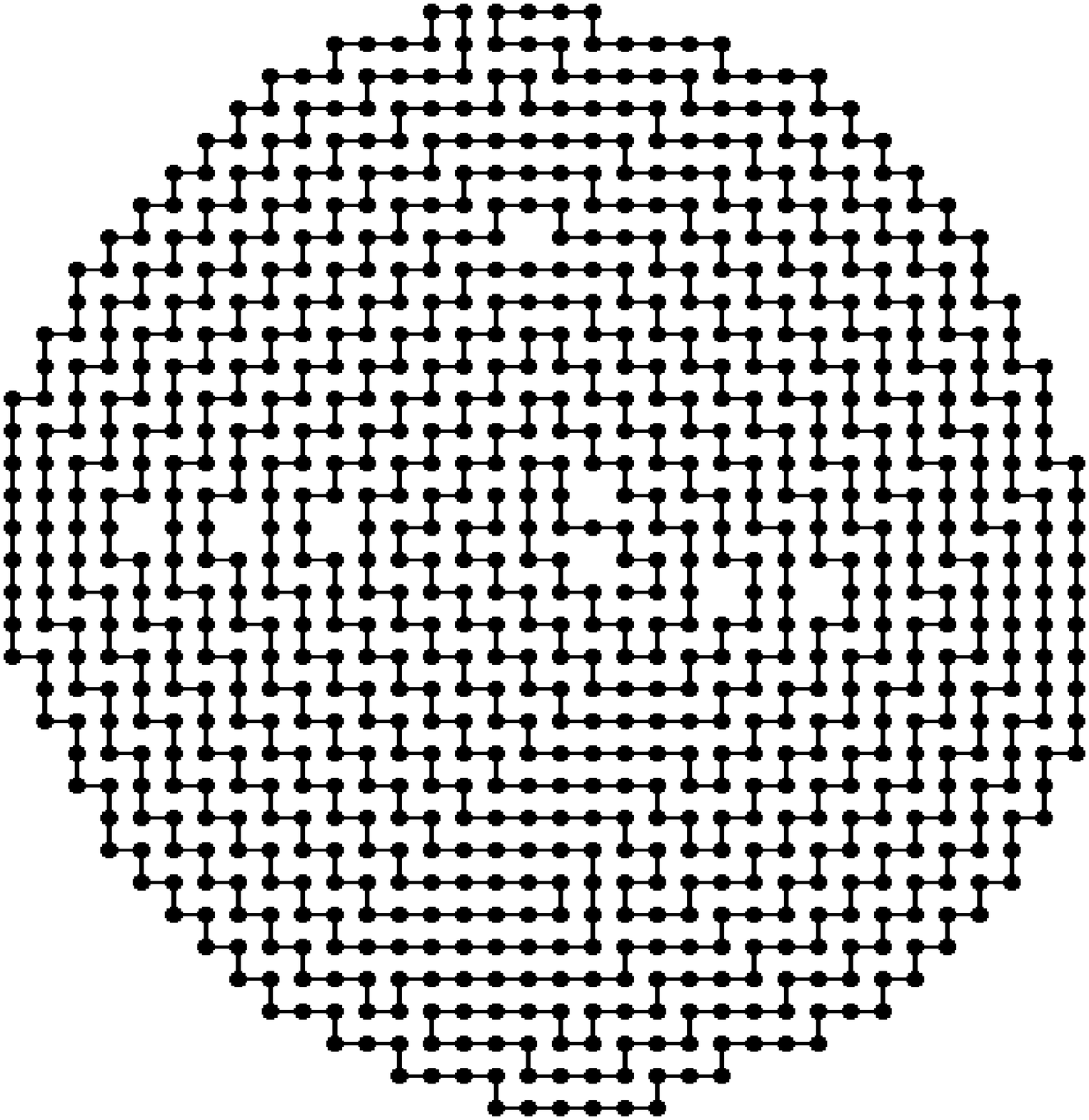}\label{grosCS}}
\caption{Obtaining unfoldable SAW constituted by two overlapped discs}
\end{figure}

The key idea leading to our first discovery of an unfoldable self-avoiding
walk is represented in Figure~\ref{pivotDunCercle}: a SAW constituted by two
subwalks that both fill almost 50\% of a disc (approximately the same radius for the two subwalks), 
when concatenated, can lead to an unfoldable self-avoiding walk if: 
\begin{itemize}
\item when superposed, they fill almost 100\% of the disc,
\item the extremities of the concatenated walks are near the center of the 
superposed disc. 
\end{itemize}
Additionally, it is required that the first subwalk ends itself by visiting its
boundary circle whereas the second subwalk starts by visiting this circle (more
precisely, an equivalent circle with a slight different radius).
These two subwalks have been obtained by following two closed spiral trajectories
in the opposite direction.

By this way, no pivot move should be realized without breaking the self-avoiding 
property. Indeed, due to the compactness of the resulted walk having the form
of a disc, no pivot move should be achieved inside the disc, whereas a pivot
move at its bounds (the circle) separates the whole disc in two overlapping 
ones, as depicted in Figure~\ref{pivotDunCercle}. A first realization of
such an unfoldable SAW is shown in Figure~\ref{grosCS}.

After obtaining our first unfoldable self-avoiding walks, the second stage
was to reduce their number of steps by removing the central part of the discs
and reducing, bit by bit, its radius. The first operation has been realized
by removing the head of the first subwalk and the tail of the second one, 
whereas the second operation consists in removing the end (resp., the beginning)
of the spirals mentioned above.
After having found a self-avoiding walk having the form of a smaller disc, the use
of the backtracking program of Listing~\ref{backtrack} has often been required
to make this SAW as unfoldable.
1840 unfoldable self-avoiding walks have been discovered by doing so, a few of them
being represented in Table~\ref{listOfSAWs}, while they are counted according to
their number of steps in Table~\ref{composante connexe}.

\begin{table}[h]
\centering
\begin{tabular}{l|l}
STEPS & RELATIVE ENCODING\\
\hline
109 & \begin{footnotesize}033222333221212121000011111122233322233333303030300000010101011111122112\end{footnotesize}\\
    & \begin{footnotesize}3330033332323232222121212111100011103\end{footnotesize}\\
111 & \begin{footnotesize}123332223333030303000010101011112221110033003333332323232222221212121111\end{footnotesize}\\
    & \begin{footnotesize}110001110003333332222303030300111000112\end{footnotesize}\\
112 & \begin{footnotesize}333222333303030300001010101111211103303333332323232222221212121111110001\end{footnotesize}\\
    & \begin{footnotesize}1100333332223303030300111101121003333011\end{footnotesize}\\
115 & \begin{footnotesize}333222333303030300001010101111211103303333332323232222221212121111110001\end{footnotesize}\\
    & \begin{footnotesize}1100333332223303030300111101122100033033223\end{footnotesize}\\
123 & \begin{footnotesize}232323233330303030000101010111121212100303033333323232322222212121211111\end{footnotesize}\\
    & \begin{footnotesize}101010100033232323233030303001010101122321232330012\end{footnotesize}\\
145 & \begin{footnotesize}222323332300033300301000111010111221122112223222232333233333030303003000\end{footnotesize}\\
    & \begin{footnotesize}0010012223222122121212111011101000010330033003232332322211121101003333321\end{footnotesize}\\    
146 & \begin{footnotesize}2223233323000333003010001110101112211221122232222323332333330303030030000\end{footnotesize}\\
    & \begin{footnotesize}0100122232221221212121110111010000103300330032323323222111211010303230032\end{footnotesize}\\
\end{tabular}
\caption{A short list of unfoldable SAWs}
\label{listOfSAWs}
\end{table}

\begin{figure}[h!]
\begin{center}
  \subfigure[$w_0$ (239-step walk)]{\includegraphics[scale=0.2]{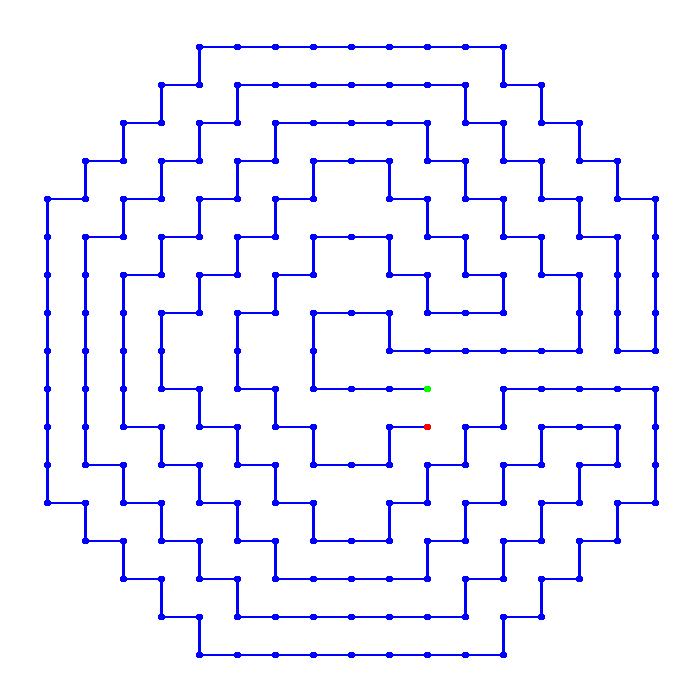}\label{nfSAW0}}\quad 
  \subfigure[$w_1$ (391-step walk)]{\includegraphics[scale=0.2]{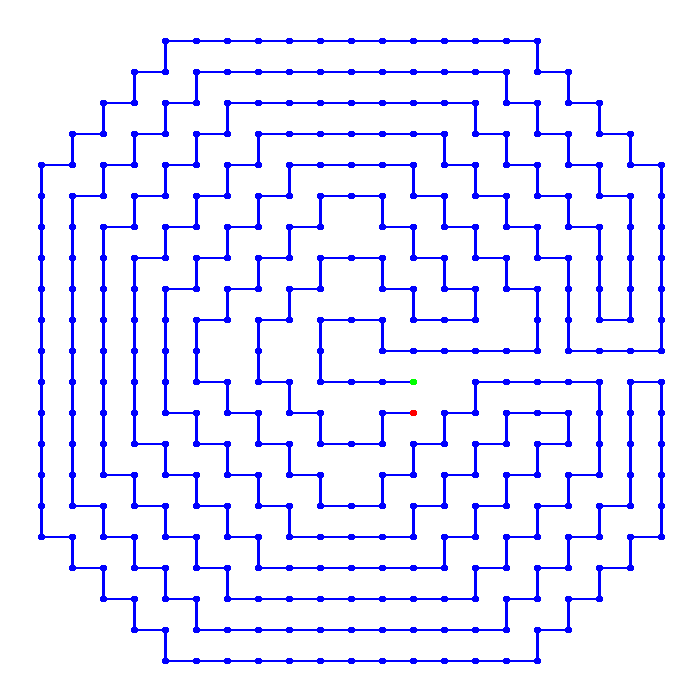}}\\
  \subfigure[$w_2$ (575-step walk)]{\includegraphics[scale=0.2]{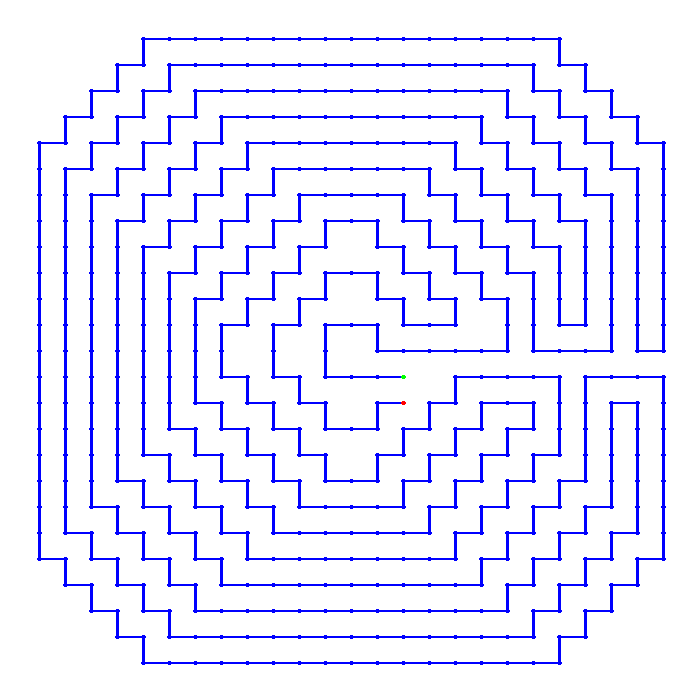}}\quad 
  \subfigure[$w_3$ (791-step walk)]{\includegraphics[scale=0.2]{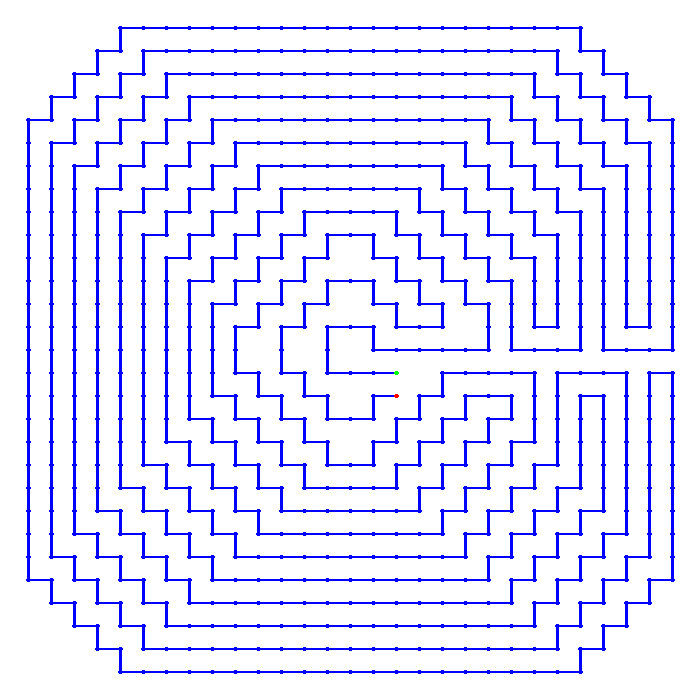}}
\end{center}
\caption{Generating walks that cannot be folded out}
\label{rtuc}
\end{figure}

A second approach to obtain unfoldable SAWs is to take a compact 
unfoldable SAW bounded by a kind of circle, and to extend it by two connected
 spirals covered in opposite direction, as depicted in Figure~\ref{rtuc} 
A proof of the correctness of this approach, leading to an infinite number of
 unfoldable self-avoiding walks, is detailed in~\cite{articleTheoreme}.
However, as by this extension-based method it is obviously impossible to obtain
the shortest unfoldable self-avoiding walk, such an approach will not be discussed
in detail in this research work (for further details about this approach, the
reader is referred to~\cite{articleTheoreme}).

\begin{table}
 \centering
 \begin{scriptsize}
 \begin{tabular}{c|c|c|c|c}
 $n$ & $\sharp\mathfrak{G}_n$ & $\sharp f^1SAW(n)$ & $\sharp USAW(n) = \sharp \overline{f^1SAW(n)}$ & $\sharp fSAW(n)$\\
\hline
1 & 4 & 4 & 0 & 4 \\
2 & 12 & 12 & 0 & 12 \\
3 & 36 & 36  & 0 &  36 \\
4 & 100 & 100 & 0 & 100 \\
5 & 284 & 284 & 0 & 284 \\
6 & 780 & 780  & 0 & 780\\
7 & 2172 & 2172 & 0 & 2172 \\
8 & 5916 & 5916  & 0 & 5916 \\
9 & 16268 & 16268 & 0 & 16268  \\
10 & 44100 & 44100 & 0 & 44100 \\
11 & 120292 & 120292 & 0 & 120292 \\
12 & 324932 & 324932 & 0 & 324932  \\
13 & 881500 & 881500 & 0 & 881500 \\
14 & 2374444 & 2374444 & 0 & 2374444 \\
15 & 6416596 & 6416596 & 0 & ? \\
16 & 17245332 & 17245332 & 0 & ? \\
17 & 46466676 & 46466676 & 0 & ?\\
18 & 124658732 & 124658732 & 0 & ?\\
19 & 335116620 & 335116620 & 0 & ?\\
20 & 897697164  & 897697164 & 0 & ?\\
21 & 2408806028  & 2408806028 & 0 & ?\\
22 & 6444560484  & 6444560484 & 0 & ?\\
23 & 17266613812  & 17266613812 & 0 & ?\\
24 & 46146397316  & 46146397316 & 0 & ?\\
25 & 123481354908  & 123481354908 & 0 & ?\\
26 & 329712786220  & 329712786220 & 0 & ?\\
27 & 881317491628  & 881317491628 & 0 & ?\\
28 & 2351378582244  & 2351378582244 & 0 & ?\\ 
29 & 6279396229332   & ? & ? & ?\\
30 & 16741957935348   & ? & ? & ?\\
31 & 44673816630956   & ? & ? & ? \\
\vdots & \vdots & \vdots & \vdots & \vdots \\
107 & ? & ? & $\geqslant 1$ & ? \\
108 & ? & ? & $\geqslant  1 $ & ? \\
111 & ? & ? & $\geqslant  5 $ & ? \\
112 & ? & ? & $\geqslant  1 $ & ? \\
113 & ? & ? & $\geqslant  2 $ & ? \\
114 & ? & ? & $\geqslant  2 $ & ? \\
115 & ? & ? & $\geqslant  5 $ & ? \\
116 & ? & ? & $\geqslant  3 $ & ? \\
117 & ? & ? & $\geqslant  4 $ & ? \\
118 & ? & ? & $\geqslant  2 $ & ? \\
119 & ? & ? & $\geqslant  2 $ & ? \\
121 & ? & ? & $\geqslant  4 $ & ? \\
122 & ? & ? & $\geqslant  5 $ & ? \\
123 & ? & ? & $\geqslant  1 $ & ? \\
132 & ? & ? & $\geqslant  7 $ & ? \\
133 & ? & ? & $\geqslant  6 $ & ? \\
134 & ? & ? & $\geqslant  95 $ & ? \\
135 & ? & ? & $\geqslant  165 $ & ? \\
136 & ? & ? & $\geqslant  40 $ & ? \\
137 & ? & ? & $\geqslant  50 $ & ? \\
138 & ? & ? & $\geqslant  175 $ & ? \\
139 & ? & ? & $\geqslant  179 $ & ? \\
140 & ? & ? & $\geqslant  66 $ & ? \\
141 & ? & ? & $\geqslant  119 $ & ? \\
142 & ? & ? & $\geqslant  322 $ & ? \\
143 & ? & ? & $\geqslant  476 $ & ? \\
144 & ? & ? & $\geqslant  8 $ & ? \\
145 & ? & ? & $\geqslant  18 $ & ? \\
146 & ? & ? & $\geqslant  54 $ & ? \\
235 & ? & ? & $\geqslant  1 $ & ? \\
239 & ? & ? & $\geqslant  1 $ & ? \\
391 & ? & ? & $\geqslant  1 $ & ? \\
575 & ? & ? & $\geqslant  1 $ & ? \\
791 & ? & ? & $\geqslant  1 $ & ? \\
 \end{tabular}
 \end{scriptsize}
\caption{Cardinalities of various subsets of SAWs}
\label{composante connexe}
\end{table}

%\begin{proof}
%Obtained experimentally, see Table~\ref{composante connexe}.
%\end{proof}

%Concerning the connected components, we have experimentaly verify that
%connected components larger than 1 or 2 elements exist. Figure shows
%an example of a 5 sized connected component.

%\begin{figure}
%  \begin{minipage}{0.3\textwidth}
%    \centering\resizebox{\textwidth}{!}{
%      \includegraphics[scale=0.15]{115_a.png}
%    }
%  \end{minipage}
%  \begin{minipage}{0.3\textwidth}
%    \centering\resizebox{\textwidth}{!}{
%      \includegraphics[scale=0.15]{115_b.png}
%    }
%  \end{minipage}
%  \begin{minipage}{0.3\textwidth}
%    \centering\resizebox{\textwidth}{!}{
%      \includegraphics[scale=0.15]{115_c.png}
%    }
%  \end{minipage}
%  \begin{minipage}{0.3\textwidth}
%    \centering\resizebox{\textwidth}{!}{
%      \includegraphics[scale=0.15]{115_d.png}
%    }
%  \end{minipage}
%  \begin{minipage}{0.3\textwidth}
%    \centering\resizebox{\textwidth}{!}{
%      \includegraphics[scale=0.15]{115_e.png}
%    }
%  \end{minipage}
%  \caption{A connected component with 5 elements}
%  \label{sawConnected}
%\end{figure}

\section{Toward Heuristic Approaches}
\label{sec:heuristics}

A first quite optimistic heuristic method to discover smaller unfoldable 
self-avoiding walks is to follow a Monte-Carlo approach,
as described in what follows. 
Starts first with the origin $(0,0)$, then at each iteration:
\begin{enumerate}
\item Pick randomly a stretching direction among the $\{0,1,2,3\}$ set.
\item If it is possible to extend the current walk in that direction while preserving
the self-avoiding property, then:
\begin{itemize}
\item do the extension;
\item if the obtained walk has the targeted number of steps, then tests if it is 
unfoldable.
\end{itemize}
\item If not, try to pick a new stretching direction again until reaching a
predefined number of attempts. If this number is reached, then restart the whole 
process.
\end{enumerate}
Of course, when picking a new direction in $\{0,1,2,3\}$, the previous one
is considered beforehand and the set is adapted: if the previous move was 0,
then the next one is picked in $\{0,1,3\}$, as direct reversals are banished
to preserve the self-avoiding property.
 
An implemented improvement of this Monte-Carlo based approach for finding
unfoldable self-avoiding walks  is to consider that such walks are perhaps
more compact than other SAWs, as any pivot move must meet the tail of the
structure. To take benefits of such an assessment, assuming that to be true,
we have required in a new version of the Monte-Carlo program that the  
walks must stay in a $N\times N$ lattice, restarting the process at each time
a SAW has an height or a width exceeding $N$.
We have experimentally limiting the lattice to a square of size $12 \times 12$
and to self-avoiding walks having a number of steps lower than 100. However,
no smaller unfoldable SAW has been discovered after 15
days of computation on the supercomputer facilities.

Remark that a second version of the backtracking algorithm has been written too,
to take into account restrictive $N \times N$ square lattices. However, this 
program has not allow the author's to discover smaller unfoldable self-avoiding
walks than the one depicted in Figure~\ref{saw107}.

\begin{figure}[h]
\centering
\includegraphics[scale=1]{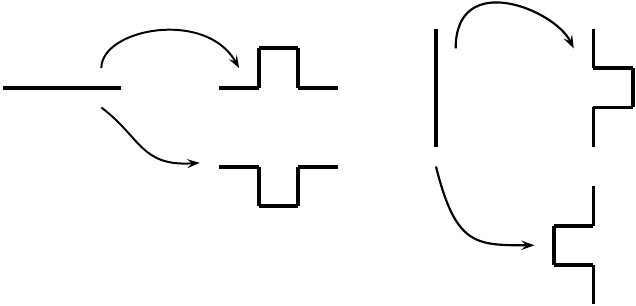}
\caption{Yet another stretching method}
\label{newMethod}
\end{figure}

Other investigated approaches encompass stretching method presented in Figure~\ref{newMethod},
the computing of the fact that the extremities of the walk should be closed one
to each other, and an adaptable probability distribution of the stretching 
direction set $\{0,1,2,3\}$ for favoring the ones that stretch the walk in the
center of the walk.
They all have lead to a failure in discovering shorter unfoldable self-avoiding
walks.

\section{PySAW software}

\begin{figure}[h]
\centering
\includegraphics[scale=0.3]{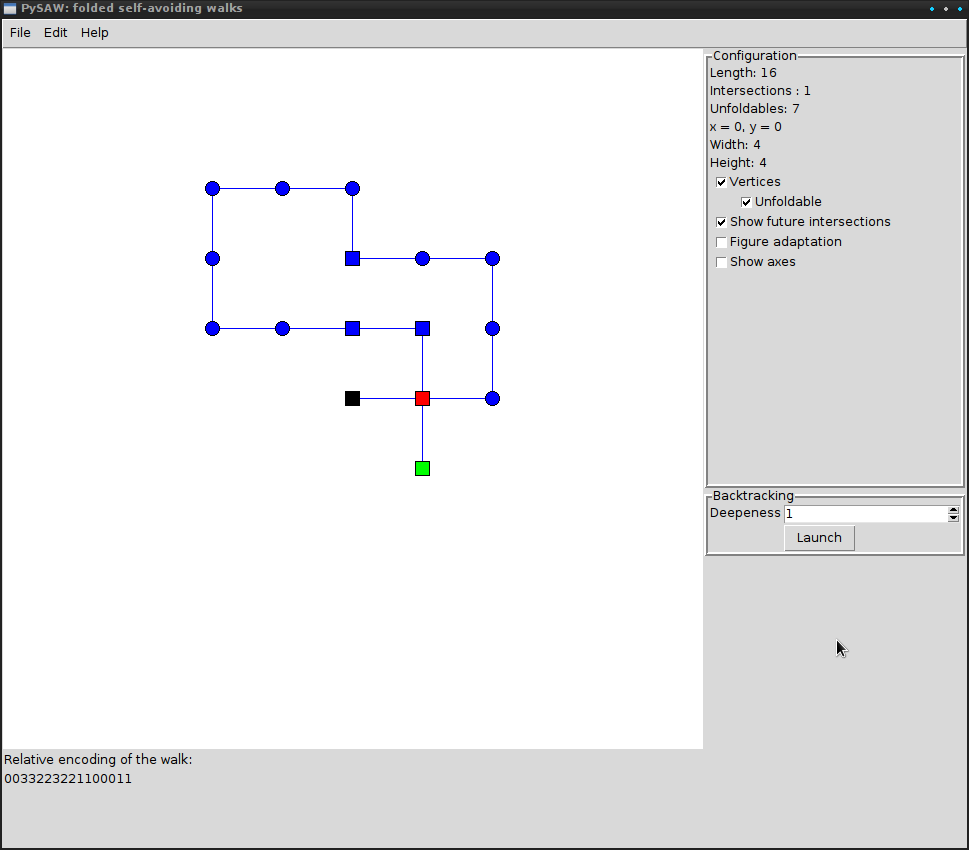}
\caption{PySAW software}
\label{PySAW}
\end{figure}

To investigate folded and unfoldable self-avoiding walks,
we have developed a Python~\cite{python} interface hosted 
in \url{https://code.google.com/p/pysaw} and freely 
available\footnote{This python code is maintained by
Christophe Guyeux (\url{christophe.guyeux@univ-fcomte.fr}, any feedback is welcome).}.
The interface is depicted in Figure~\ref{PySAW} and its 
current functionality is detailed below.

The interface is divided in four parts: a sandbox where the SAW
is drawn, two frames for configuration and backtracking, and
the relative encoding of the depicted self-avoiding walk at the
bottom of the window. A menu bar with common File, Edit, and Help
items completes the interface.

In the sandbox frame, is is possible to construct a SAW step 
by step by using arrow keys (left, right, up, bottom). The initial
vertex is the black square while the green square represents the
last vertex. Other vertices and the remainder of the walk are
 represented in blue. A vertex $V$ is represented by a circle 
 when at least one of the two $\pm 90^{\circ}$ pivot moves on
 $V$ leads to a new walk satisfying the self-avoiding property,
 whereas it is a square when this vertex is unfoldable.
Intersections in the walk, for its part, are depicted by a red square,
as represented in Figure~\ref{PySAW}.

Various modification capabilities of the depicted walk have been 
added. Del key removes the last vertex whereas Back Space 
one deletes the first vertex. An item in the Edit menu allows the
deletion of a subwalk between two vertices selected by the right
mouse button: the two remainder subwalks are then concatenated.
The Edit menu contains too an item that reverse the walk (maps
$w(0)...w(n)$ in $w(n)...w(0)$.
A click of the left mouse button on a vertex realizes 
a $+90^{\circ}$ pivot moves on this point whereas a right
click makes a pivot moves in the reverse direction.
Another computed facility is the capability to extend a walk
between two points by a right button drag-and-drop: the walk
between the two extremities (vertices) of this drag-and-drop
is replaced by the walk of the mouse motion.
Control-s is a fast save of the walk in \emph{length.txt} file,
while Control-q quit the program.
Finally, Control-z combination undoes the last modification whereas
Control-Z redoes it.
Let us remark that, at each modification, the relative encoding 
of the walk provided at the bottom of the window is naturally
updated.

\begin{figure}[h]
\centering
\includegraphics[scale=0.3]{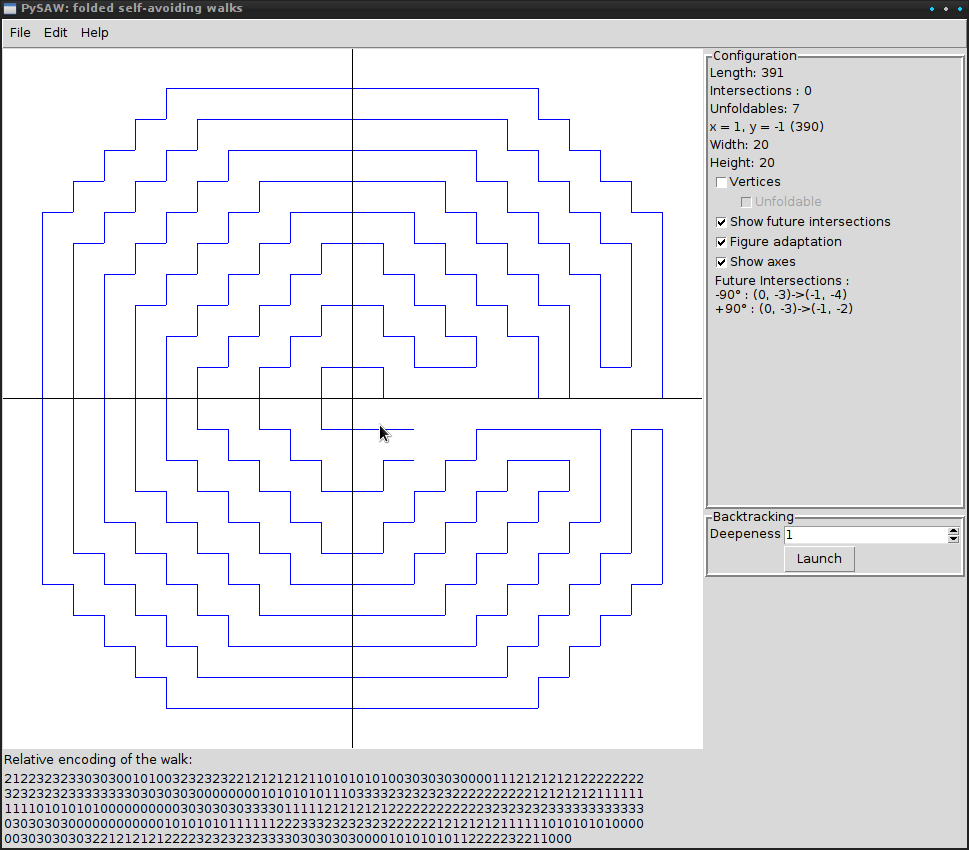}
\caption{PySAW software}
\label{PySAW2}
\end{figure}

The representation of the walk in the sandbox can be altered by two check boxes in
the Configuration panel. The Vertices check box enables or not
the representation of the vertices as circles. When
enabled, a second check box entitled ``Unfoldable'' enables the
computation of each vertex: unfoldable (square) or not (circle).
In Figure~\ref{PySAW} this check box is enabled whereas it is
disabled in Figure~\ref{PySAW2}.
To disable this check box is a necessity for very long walks, due
to computation time.
Another check box specifies if origin and axes must be
represented in the sandbox
(the origin can be changed in the Edit menu).
Other information represented in the Configuration panel are the
number of steps of the walk (length), its width and height, 
its number of intersections and of unfoldable vertices.
Such information is used when representing the walk. 
Indeed, when the check box ``Figure adaptation'' is enabled, 
then the representation of the walk is modified in such a way that
this walk is always contained in the frame: the length of the step
and the position of the walk are adapted to the frame, 
depending on the weight
and height of the walk. If not enabled, the walk can goes outside the 
frame during a pivot move (for instance), which can be desired to 
understand well the effects of a given pivot move: the head of the
walk does not move, is not adapted to the figure, during this 
pivot. 

The position of the mouse is given too in the Configuration panel,
depending on te chosen origin. Initially, this origin is set to
the first node of the walk but, as stated before, it is possible
to change it using New origin of the Edit menu. When the mouse
is positioned on a vertex $w(k)$ of the walk, its position $k$
is informed in parenthesis and, if the ``Show future intersection''
check box is enabled, the list of intersections implied by 
$\pm 90^{\circ}$ pivot moves on $w(k)$ is provided at the bottom of
the Configuration panel.

Other functionality of the PyUSAW software encompasses a backtracking
search of unfoldable SAWs, which is based on Listing~\ref{backtrack}: the self-avoiding walk of the sandbox is 
extended systematically until reaching the specified deepness and,
at each time an unfoldable SAW is found, its relative encoding is
print in the terminal. Drawn self-avoiding walks can be
saved as a relative encoding text file, which can contain or not 
the position of $w(0)$ in the square lattice, and naturally these
text files can be opened by PyUSAW. Finally, the walk can be 
exported too in various image formats (SAWs represented in this
research work have been obtained using this interface).

\section{Conclusion}

In this article, various computational methods have been proposed
to investigate the newly discovered folded and unfoldable subsets
of walks. These methods encompass backtracking, breadth first
search, and Monte-Carlo approaches. The targeted goals
was to find when the set of folded SAWs becomes different from the
set of all self-avoiding walks, and to discover the shortest 
unfoldable SAWs. Significant advances have been achieved and 
explained with detail, and an original Python tool to facilitate
the study of these important SAW subsets has finally been presented.

In future work, the authors' intention is to restrict the range
of uncertainties regarding the smallest $n$ such that 
$fSAW(n) \neq \mathfrak{G}_n$, which is currently known to belong
in $\llbracket 15, 107\rrbracket$. To do so, computational methods
presented in this article, often quite naive, will be enhanced and
optimized. Heuristic approaches encompassing swarm particles and
genetic algorithms, will be regarded too to discover shorter 
unfoldable SAWs, if exist. Theoretically speaking, a complexity 
study of the protein structure prediction problem in the subset
of folded SAWs will be realized, and we will try to 
rewrite the generating function of folded SAWs in other generating
functions easier to calculate. Finally, consequences regarding
the best ways to make protein structure prediction will be 
investigate.

\section*{Acknowledgement}
The authors wish to thank Thibaut Cholley, Raphaël Couturier, Jean-Marc Nicod, 
and Alain Giorgetti for their help in understanding folded and unfoldable SAWs. All
the computations presented in the paper have been performed on the
supercomputer facilities of the Mésocentre de calcul de Franche-Comté.

\bibliographystyle{plain}
\bibliography{biblio}

\end{document}